\documentclass[11pt]{article}
\usepackage{moriond2000,epsfig}

\def\Journal#1#2#3#4{{#1} {\bf #2}, #3 (#4)}

\def\APJ{\em ApJ}

\def\be{\begin{equation}}
\def\ee{\end{equation}}
\def\bea{\begin{eqnarray}}
\def\eea{\end{eqnarray}}

\def\simlt{\lower.5ex\hbox{$\; \buildrel < \over \sim \;$}}
\def\simgt{\lower.5ex\hbox{$\; \buildrel > \over \sim \;$}}
\def\mag{\mbox{ mag}}

\def\kpc{\mbox{ kpc}}

\def\msun{\mbox{ M}_\odot}

\begin{document}
\vspace*{4cm}
\title{PROSPECTS FOR THE LENSING OF SUPERNOVAE}

\author{ R. BENTON METCALF }

\address{Institute of Astronomy, Cambridge \\ Madingley Road\\
Cambridge CB3 0HA, England}

\maketitle
\abstracts{
Observations of high redshift type Ia supernovae (SNe)
will enable us to probe the structure of galaxy halos and the
composition of dark matter.  The future prospects for this field are
briefly discussed here.  First the ability of SN observations to
differentiate between dark matter made of macroscopic compact objects
and dark matter made of microscopic particles is reviewed.  Then a new
method for probing the structure of galaxy halos and galaxy cluster
halos is described.  This method utilizes the correlations between
foreground galaxy light and supernova brightnesses to substantially
decrease possible systematic errors.  The technique may be particularly
useful for measuring the size of dark matter halos, a measurement to
which the galaxy--galaxy lensing is not well suited, and the level of
substructure in galaxy halos, a problematic prediction of the cold dark
matter model.  The required observations of hundreds of SNe at $z \sim
1$ are already being proposed for the purposes of cosmological parameter
estimation.
}

\section{Introduction}

There are still many outstanding problems in the fields of structure
formation and dark matter which gravitational lensing has not yet been
able to address.
Outside of the visible extent of galaxies not a great deal is known
about the distribution of dark matter (DM) on scales smaller than galaxy
clusters.  Galaxy--galaxy lensing has put some constraints on
the mass of galaxy halos, but their size scale and the degree to which
they are smooth mass distributions or collections of subclumps are not
well determined.\cite{astro-ph/9912119,1998ApJ...503..531H,1996ApJ...466..623B,1996MNRAS.282.1159G,1984ApJ...281L..59T}  Even the
concept of galaxy halos, with a single galaxy within each of them and
well defined sizes, may not be correct.  Nor is it clear how the observable
properties of galaxies relate to the DM distribution around them.  Recently dark
matter simulations have revealed several problems with the cold dark
matter (CDM) model.  One of these is that CDM predicts a large amount 
of small scale structure within halos which is unaccounted for in the
observed distribution of light.\cite{1999ApJ...524L..19M}  In addition,
the composition of DM is still a mystery: it could be large compact
objects like black holes or it could be microscopic particles like
WIMPS.  Even with primordial nucleosynthesis bounds roughly half the
baryons are in some form that has yet to be directly detected --
conceivably in condensed objects.  I will describe here how the
gravitational lensing of type Ia SNe can help to answer some of these
questions.  Lack of space dictates that this only be an outline and that
the reader be referred to more detailed papers.

High redshift type Ia SNe have recently been used by two collaborations
to measure cosmological
parameters.\cite{1997ApJ...483..565P,1998AJ....116.1009R} 
These measurements and the proposed application of Ia's to gravitational
lensing are made possible by the discovery of a tight correlation
between the
light curve shapes and peak luminosities of these SNe.\cite{hamuy96,RPK96}
The measured standard deviation of corrected peak luminosities in local
SNe is now $\simlt 0.12 \mag$.  The
successes of these collaborations have inspired plans for more aggressive,
larger searches for high redshift SNe in the near future.  The volume and
quality of data is likely to increase dramatically.
Besides constraining $\Omega_m$ and $\Omega_\Lambda$ these data, amongst
other things, would be ideal for gravitational lensing studies.

In the next section I will discuss how SNe can be used to determine if
dark matter is composed of macroscopic compact objects and in
section~\ref{corrilations} the case of DM composed of microscopic
particles is taken up.

\section{Lensing by macroscopic compact objects}\label{compact}

\begin{figure}[t]
\begin{center}\epsfig{figure=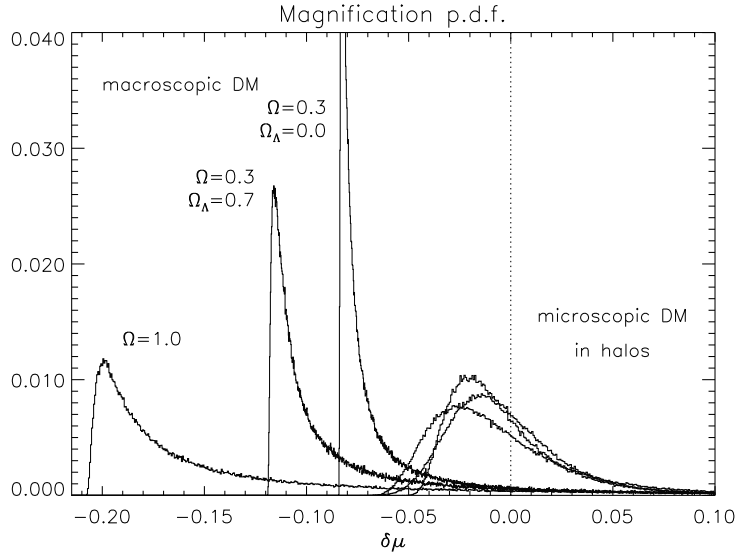,height=3in}\end{center}
\caption{These are the magnification probability distributions for point
sources in several models.  The three distributions peaking on the left
are for DM composed entirely of macroscopic compact objects.  The
background cosmologies are listed.  The more centrally concentrated
distributions on the right are for microscopic DM concentrated into
galaxy halos.  These distributions are functions of both the background
cosmology and the halo model.  Changing the halo model makes relatively 
little difference if the fraction of mass in halos is kept fixed.
\label{fig:pdf}}
\end{figure}
\begin{figure}[t]
\begin{center}\epsfig{figure=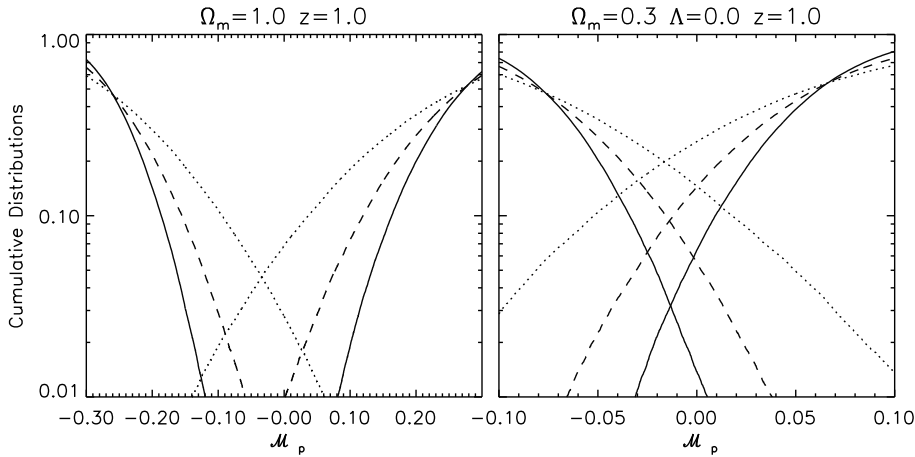,height=4in}\end{center}
\vspace{-0.75in}
\caption{The cumulative distributions of ${\mathcal M}_p$
for different models and different numbers of observed SNe.  The dotted
curves are for 21 observed SNe, the dashed for 51 and the solid for
101.  The pairs of like curves represent the cases of macroscopic and
microscopic DM.  The distributions for the cases of
macroscopic DM rise to the right while if DM is microscopic ${\mathcal
M}_p$ is expected to be smaller.  The overlap of these distributions is
small signifying that DM candidates could be distinguished using this
statistic.  For a flat cosmological model the constraints are stronger
than for the $\Lambda=0$ model with the same $\Omega_m$.  The variance in
SN luminosities is $\sigma_{sn}=0.16\mag$.
\label{fig:M_p}}
\end{figure}
Some magnification probability distributions for a point source at $z=1$
are plotted in figure~\ref{fig:pdf}.  The means of all these distributions
are zero which corresponds to the usual Friedman-Robertson-Walker (FRW)
luminosity distance.  When DM is made entirely of macroscopic compact
objects the distribution peaks well below the mean making most SNe
under-luminous. 
The peak of the distribution is just slightly brighter than the
solution corresponding to the empty beam or Dyer--Roeder luminosity distance
\cite{1974ApJ...189..167D}.  This is the formal solution of Sachs optical
scalar equations\cite{sachs61} for a beam that passes through only empty
space (no Ricci focusing) and has no shear on it.  The long, high
magnification tail to these distributions correspond to rare cases where
a DM object is very close to the line of sight.  At $z<2$ it is unlikely
that multiple lens lie so close to the line of sight that their lensing
effects become nonlinearly coupled which significantly simplifies
calculations.

The distributions for microscopic DM shown in figure~\ref{fig:pdf}
assume that the DM is clustered into halos surrounding galaxies.  The
exact form of the clustering is not important for this section.  As the
figure shows, these
distributions are significantly more centrally concentrated about the
FRW solution.  
With macroscopic DM the magnification is dominated by Ricci focusing --
the isotropic expansion or contraction of the image caused by matter
within the beam -- while in the macroscopic case (with true point
sources) it is entirely due to shear caused by matter outside of the
beam.  This is the essential difference between the two cases and why
they can be differentiated using SNe observations.

To differentiate between DM candidates we construct the statistic 
\begin{eqnarray}
{\mathcal M}_p\equiv \frac{1}{N_{sn}} \ln\left[\frac
{ P\left(\{\delta\mu\} | \mbox{ macro DM,noise}\right)}
{ P\left(\{\delta\mu\} | \mbox{ micro DM,noise}\right)}\right].
\label{Mp}
\end{eqnarray}
where the $P$'s are the probability of getting the $N_{sn}$ observed
SN magnifications given that DM is of the specified type and given the
expected noise.  ${\mathcal M}_p$ is close to normally distributed for a
modest number of SNe.  Some example distributions of ${\mathcal M}_p$
are plotted in figure~\ref{fig:M_p}.  If DM is microscopic ${\mathcal
M}_p$ is expected to be small while if DM is macroscopic it will be large.
Figure~\ref{fig:M_p} shows that with 100 SNe at $z\sim 1$ one is
unlikely to confuse the two cases.

This technique works for DM objects with masses $\simgt 10^{-3}\msun$.
For smaller masses it is probable that the expanding SN photosphere
will make the magnification time dependent which is an interesting
subject in itself.  For more details on differentiating DM candidates
see Metcalf \& Silk.\cite{1999ApJ...519L...1M}  

\section{Correlations between light and magnification}\label{corrilations}

\begin{figure}[t]
\begin{center}\epsfig{figure=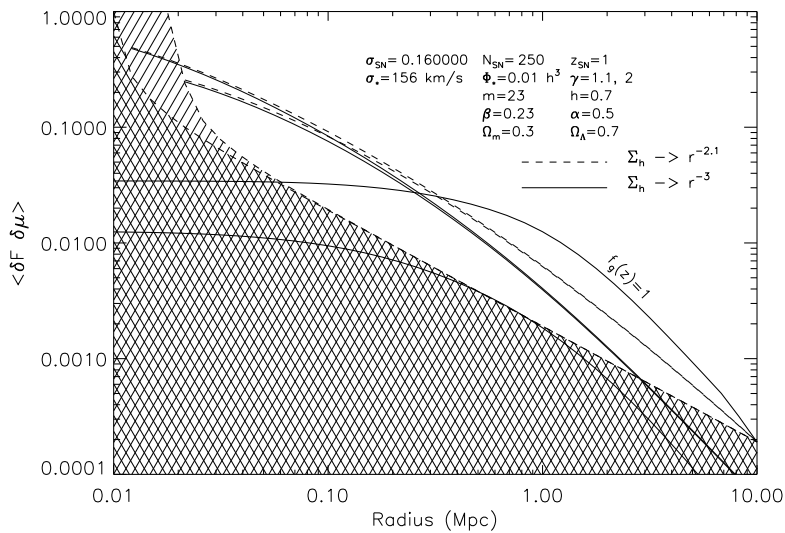,height=3.0in}\end{center}
\caption{The expected correlation between foreground light and SN
luminosity.  There are two contributions plotted separately here.
The first contribution is from {\it galactic halos} which are presumed
to surround every galaxy and have properties which are related to the
observed properties of their resident galaxy.  In addition there is a
contribution from {\it extragalactic halos} or galaxy clusters which
contain multiple galactic halos and additional matter within them.
The total correlation is the sum of the two components.
The galactic halo curves are the steeper ones starting in the upper left
corner.  Two different models which
differ in the logarithmic slope of the surface density at large
galactic radii are shown.  At small galactic radii the galactic halos
approach the form of singular isothermal spheres.  For each model two inner cutoffs are
considered -- $R_{min}=10, 20\kpc$.  To avoid obscuration all SNe within
$R_{min}$ of any galaxy are excluded.  The two curves that flatten out
on the left are the contributions from extragalactic halos in the CDM
model.  These halos have Navarro, Frenk and White profiles.  The one
marked $f_g(z)=1$ represents the case where only SNe behind
$M>10^{14}\msun$ clusters are selected.  The crisscrossed region
represents the noise contributed by uncertainties in SN peak
luminosities ($\sigma_{sn}=0.16\mag$) and shot noise in the foreground
galaxy counts for the case of 250 observed SNe.  The signal to noise ratio at
$R=200\kpc$ is $\simeq 4\sqrt{N_{sn}/250}$.  The limiting magnitude of
foreground galaxies is $m=23$.
\label{fig:S_p}}
\end{figure}
When the DM is made of microscopic particles it can be treated as a
transparent, massive fluid clumped into structures that are much larger
than the beam size.  Because of lensing the variance of SN brightnesses
will increase with redshift\cite{1999MNRAS.305..746M} which introduces
noise into the cosmological parameter estimates.  The increases in
variance is not the best way of detecting the lensing however.  Using the
correlation between foreground light and SN brightness reduces possible
systematic errors associated with the evolution of the SN population
and makes the measurement more directly sensitive to the mass and size
scale of dark matter halos.

Lets define the weighted foreground flux for each SN as
\begin{eqnarray}
{\mathcal F} = \sum_{y_i< R} w(z_i,z_s) f_i
\end{eqnarray}
where $f_i$ the observed flux from the $i$th galaxy and $y_i$ is its
distance (angular or proper) from the line of sight to the SN.  The weight
function depends on the galaxy and SN redshifts.  Figure~\ref{fig:S_p}
shows the expected 
correlation of ${\mathcal F}$ with SN brightness, $\langle \delta
{\mathcal F} \delta b \rangle$, for a SN at $z_s=1$.  This quantity is
proportional to the average surface density within distance $R$ of a galaxy.

By measuring this correlation the size scale of galactic halos could be
constrained significantly better than with galaxy--galaxy lensing using
a much larger data set.  This is a result of the magnification being a
steeper function of galactic radius than shear when the density profile
is steeper than isothermal.  In addition, by selecting or searching for
SNe behind galaxy clusters the structure of clusters and the smaller
halos within them can be probed.  For instance the tidal truncation of
galactic halos could be investigated.  Higher order correlations can also
be used to look for substructure in galaxy halos.  This subject is
treated more thoroughly in a complete paper by the author.\cite{Metcalf00}

\section{Discussion}

The number of high redshift SNe required for the studies proposed here
are well within the projected numbers for future SN searches -- the VISTA 
telescope is expected to find hundreds at $z \simgt 1$.\footnote{home
page: http://www-star.qmw.ac.uk/$\sim$jpe/vista/} and the
proposed SNAPSAT satellite could find thousands up to $z \simeq
1.7$.\footnote{home page: http://snap.lbl.gov/proposal/}  The errors
assumed in this paper are quite conservative compared to those expected
for satellite observations.  The future looks bright for this new field
in gravitational lensing.

\section*{Acknowledgments}
Many thanks to all the friends that made this such an enjoyable conference.  
Thanks also to J. Silk for all his help while some of this work was being done.

\end{document}